\documentclass[twocolumn,showpacs,preprintnumbers,amsmath,amssymb]{revtex4}

\usepackage{graphicx}
\usepackage{dcolumn}
\usepackage{bm}
\usepackage{color}

\begin{document}


\title{Muon Acceleration in Cosmic-ray Sources}
\author{Spencer R. Klein} \altaffiliation{Lawrence
Berkeley National Laboratory, Berkeley CA 94720 USA \& Physics Dept. University of California, Berkeley, CA 94720 USA.} 
\author{Rune Mikkelsen} \altaffiliation{Lawrence Berkeley National
 Laboratory, Berkeley CA, 94720, USA \& Department of Physics and Astronomy, Aarhus University, DK-8000 Aarhus C, Denmark}
\author{Julia K.\ Becker Tjus} \altaffiliation{Fak.\ f.\ Phys.\ \& Astron., Ruhr-Universit\"at Bochum, Germany}
 

\begin{abstract}
Many models of ultra-high energy cosmic-ray production involve acceleration in linear accelerators located in Gamma-Ray Bursts magnetars, or other sources.   These source models require very high accelerating gradients, $10^{13}$~keV/cm, with the minimum gradient set by the length of the source. 
At gradients above
1.6 keV/cm, muons produced by hadronic interactions undergo significant acceleration before they decay.  This acceleration hardens the neutrino energy spectrum and greatly increases the high-energy neutrino flux.  We rule out many models of linear acceleration, setting
strong constraints on plasma wakefield accelerators and on 
models for sources like Gamma Ray Bursts and magnetars.

\end{abstract}

\maketitle

For more than 100 years, scientists have investigated cosmic-rays (CRs)
and their origin. 
Up to energies of $\sim 5$ PeV, CR production
may be adequately described by current models of acceleration in
supernova remnants. 
Many sources have been proposed as the acceleration sites for
ultra-high energy (UHE) CRs, including
Active Galactic Nuclei (AGNs), Gamma Ray Bursts (GRBs), and
magnetars. Typically, acceleration
involves repeated diffusive shock acceleration, at a site where a magnetic field
confines the particles \cite{fermi1949,bell1978,schlickeiser1989,meli2008}.

Other authors have proposed single-pass linear accelerators
\cite{Historical1,Historical2,Wakefield,Arons2003,MagnetoWakefield,Kotera,
  Pulsars}, often in a relativistic jet, such as those emerging
from GRBs or AGNs. Here, the maximum CR energy and the finite
source length imposes a lower limit on the accelerating gradient.
Further, the short time scales associated with possible transient
sources, such as GRBs and magnetars, argue for very compact
accelerators and thus requires large gradients. 

The needed accelerating gradients may be provided by a couple of mechanisms.  
Plasma wakefield acceleration (PWA) allows for gradients up to $\approx 10^{16}$ eV/cm. PWA may be present at a variety of accelerating sites, including GRBs and AGNs.    Magnetars, newborn neutron stars with enormous (petagauss) magnetic fields, also allow single-pass acceleration with very high gradients \cite{Arons2003}.  

One method to locate UHE acceleration sites is to search for neutrinos which point back
to their production site.
The neutrinos come from the decay of pions and kaons which are produced when CRs interact with nuclei or photons in the source.   Neutrinos from $\pi^\pm$ or K$^\pm$  are produced in a ratio $\nu_\mu:\nu_e:\nu_\tau = 2:1:0$,
but oscillate to $\sim 1:1:1$ before they reach Earth \cite{review2008}.  
Prompt neutrino production is mainly a factor at very high energies, above 1 PeV for current gamma-ray bursts \cite{Enberg:2008te}. 

Calculations of neutrino production in CR accelerators use the measured CR flux 
and assumptions about the CR composition and nuclei/photon target density in the source to predict a neutrino flux that might be seen in terrestrial detectors. 
The neutrino flux can be affected by energy loss or acceleration of muons.  
Muon energy loss is expected to be dominated by
synchrotron radiation.  For protons, 
this may be important at energies above $10^{17}$ eV
\cite{KashtiWaxman,Winter}.  
At large accelerating gradients, muons will gain energy before they decay, altering the
neutrino energy spectrum and composition. This was considered for choked GRBs in \cite{KoersWijers}. When the energy gain is large enough, the neutrino flux will be enhanced significantly.

In this letter, we study the effect of muon acceleration and energy loss on neutrino production, and determine the limiting energies and the enhanced fluxes.  We use existing neutrino flux limits to set limits on the maximum accelerating gradients, under the assumption that these sources are the sites of CR acceleration, and show that linear acceleration models are strongly constrained by current
neutrino flux limits.

The muon energy gain in a distance $d$ at a gradient $g$ is $\Delta E = gd$.   
For small gradients,  $g < g_0 = m_\mu c/\tau_\mu\approx 1.6$ keV/cm, 
the fractional increase in energy is small.
At higher gradients, with the relativistic increase in lifetime, the
average final energy $E_f$ of an ultra-relativistic muon is a multiple of  the initial energy $E_i$: $E_f = E_i \exp^{g/g_0}$.
This is for the mean lifetime; longer-lived muons are accelerated more,
and contribute disproportionately to the $\nu$ flux.  High acceleration turns an initially mono-energetic muon beam into a power law spectrum, with exponent $-g_0/g$.  This also applies to pions, with the appropriate mass and
lifetime substitutions.  Because the $\pi^\pm$ lifetime $\tau_\pi$
is about 1\% of $\tau_\mu$, pion acceleration is only relevant at
100 times higher gradients. 

Muons lose energy by synchrotron radiation and by interacting
with matter or photons.  Muon interactions with matter are far weaker
than for protons, with a typical range more than 100 times higher than
the proton interaction length.  So, we neglect muon interactions with
matter.

In photon-dominated regions, muons lose energy by pair conversion and
inverse Compton scattering.  These losses  are usually smaller than
those experienced by protons, which predominantly lose energy by
photoexcitation to a $\Delta$ resonance.  

We assume that, in any environment where protons are accelerated, muon energy loss is dominated by synchrotron radiation.  The energy loss in a magnetic field $B$ 
by a muon with Lorentz boost $\Gamma$ is
\begin{equation}
\frac{dE}{dt} = \frac{4}{9}c \Gamma^2B^2 \left(\frac{1}{4\pi\epsilon_0}\frac{e^2}{m_\mu c^2} \right)^2 
\end{equation}
where $\epsilon_0$ is the permittivity of free space.
For a given energy, muons have a Lorentz boost 10 times higher
than protons, so losses are 100 times higher for muons than protons at the same energy. 

\begin{figure}[htbp]
\begin{center}
\includegraphics[width=0.5\textwidth]{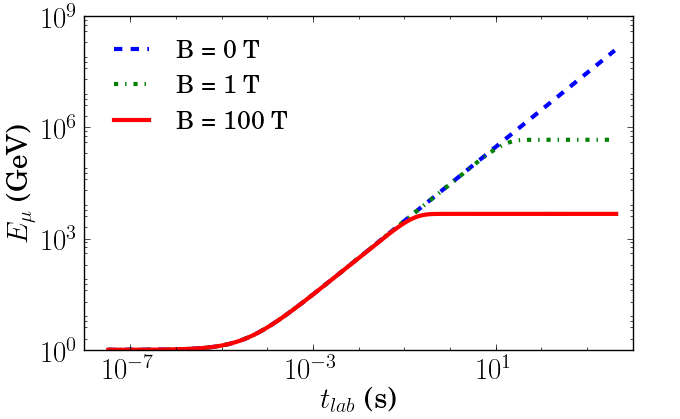}
\caption{Muon energy vs. time for acceleration in a gradient of g = 1 keV/cm, for various magnetic field strengths.}
\label{fig:EnergyOfTime}
\end{center}
\end{figure}

In the presence of acceleration and energy loss, muons reach an equilibrium energy $E_c$ at which energy gain and loss are equal. For synchrotron radiation, this energy is 
\begin{equation}
E_c = 6 \pi \frac{\epsilon_0(m_\mu c^2)^2}{B e^2}\sqrt{g}. 
\end{equation}
  Figure \ref{fig:EnergyOfTime} shows how particles gain energy at a constant rate until they reach $E_c$.
Any accelerator that can produce
$10^{20}$ eV protons can also support $10^{18}$ eV pions and muons.  This is a minimum constraint; if energy loss is dominated by interactions with matter or photons, then the 
$E_{\mu,{\rm max}}/E_{p,{\rm max}}$ will be higher.

From the decay of the charged pions, three neutrinos are produced, $\pi \rightarrow \mu\,\nu_\mu \rightarrow e\,\nu_e\,\nu_\mu\,(\nu_\mu)$; we assume that the initial pion
energy is divided equally between the three neutrinos produced and
the final state electron 
\cite{review2008}.  Muon acceleration or deceleration alters the energy of the two neutrinos coming from the muon decay, while pion acceleration or deceleration affects all of the final state particles. 

We assume that protons are accelerated following an $E^{-2}$ spectrum, up to a cutoff energy of $10^{20}$ eV.  Diffusion steepens the spectrum, leading to the observed CR spectrum, which  follows $E^{-2.7}$ ($E^{-3}$ at energies above a few PeV).  During acceleration, CRs interact with matter or photons.  We neglect the energy dependence of this cross-section (including the threshold for photopion production), so the produced pions also follow an $E^{-2}$ spectrum.  From that initial pion spectrum, we determine the energy spectrum when they decay, allowing for energy gain or loss.
The resulting muons are then followed in the same manner.   

\begin{figure}[htbp]
\begin{center}
\includegraphics[width=0.5\textwidth]{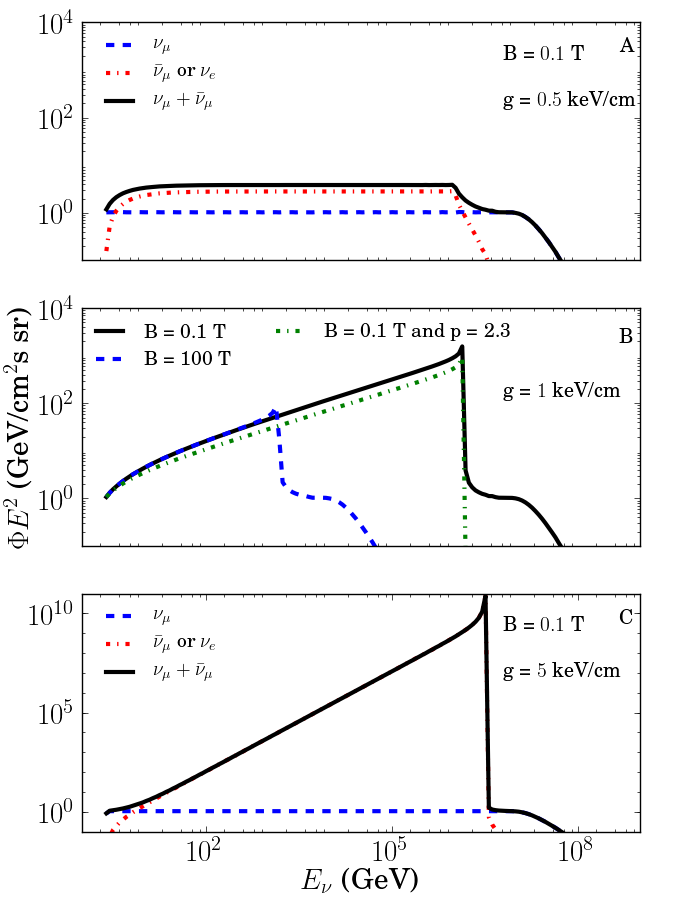}
\caption{(a) The flux of different neutrino flavors at the source for a 0.5 keV/cm gradient.   Even at this low gradient, the $\nu$ flux roughly triples.
(b) The all-flavor flux for a 1 keV/cm gradient, for different magnetic fields.  As the magnetic field rises, $E_c$ drops.  Also shown is the $\nu$ spectrum for an initial $E^{-2.3}$ proton spectrum.  
(c) The $\nu$ flux for an accelerating gradient of 5 keV/cm.  The neutrino
flux increases by many orders of magnitude.}
\label{fig:nuflux}
\end{center}
\end{figure}

Figure \ref{fig:nuflux} shows the neutrino spectrum for three accelerating gradients, including energy loss due to synchrotron radiation, compared to an un-enhanced $E^-2$ spectrum.  For gradients $g> g_0$,  the $\overline\nu_\mu$ and $\nu_e$  flux is greatly increased; even for $g$=0.5 keV/cm the flux triples.  For $g \gg g_0$,  an initially $E^{-2}$ spectrum hardens significantly, and then cuts off at $E_c$.   The flux of neutrinos with very low energies is depleted because they are accelerated to higher energies.   
As Fig. \ref{fig:nuflux}c shows, at slightly higher gradients, 5 keV/cm, the spectrum hardens (to roughly a $-g_0/g$ power law) and the
flux increases by orders of magnitude,  up to energy $E_c$.  Panel b shows the
spectrum from an initially $E^{-2.3}$ proton spectrum, at a 1 keV/cm gradient; the effect of acceleration is similar.  At higher gradients, the initially $E^{-2.3}$ spectrum is hardened to roughly $E^{-1.3}$, up to the cutoff energy of $E_c$.  

For these gradients, pion acceleration is negligible.  At much larger gradients, $g> m_\pi c/\tau_\pi\approx$ 160 keV/cm, pion acceleration would be important, and the $\nu_\mu$ curve would show similar alterations.

Neutrino production in UHE CR sources can occur in proton-proton or proton-photon interactions.   Some $\nu$ flux calculations are based on specific source models, while others are more generic, tied to the measured UHE cosmic-ray flux \cite{Waxman-Bahcall}, or the TeV photon fluxes observed by air Cherenkov telescopes.  Current diffuse $\nu$ flux limits are near or below the more optimistic flux predictions \cite{ICDiffuse,Baikal,Antares}.    

We use the IceCube diffuse $\nu$ flux limits to set two-dimensional
limits on the source opacity and accelerating gradients.   The source
opacity, $O$, treatment follows Waxman and Bahcall
\cite{Waxman-Bahcall}, who related neutrino production in an
accelerator to the matter/photon density in the accelerator. Opacity
is defined as the fraction of proton energy that is lost due to
proton-proton or proton-photon interactions in the target.  Waxman and Bahcall  set an upper limit $O=1$, corresponding to a neutrino flux limit for an $E^{-2}$ spectrum, of $\phi_{WB} = E^2\phi < 2.0\times 10^{-8}$ GeV/(cm$^2$s
sr) \footnote{This flux has been repeatedly adjusted in the literature for
different neutrino flavors or initial conditions.  For consistency with the IceCube studies, we use the same value as in Fig. 10 of \cite{ICDiffuse}.}, beyond which protons will not emerge from the acceleration site. 

We use the IceCube 40-string $\nu_\mu$ flux  limit \cite{ICDiffuse}.  For an $E^{-2}$ spectrum, the 90\% confidence level limit is $E_\nu^2\phi_{IC} = \phi < 8.9\times 10^{-9}$ GeV/(cm$^2$s sr); for an $E^{-2}$ spectrum,  90\% of the detected
events will be in the energy range between 35 TeV and 7 PeV.   This energy range is comfortably below the $10^{18}$ eV 
minimum $E_c$ for an accelerator that can produce $10^{20}$ eV protons. 

Because of the changing $\nu$ spectral indices, one cannot directly use the IceCube limit.  However,  Fig. 5 of Ref. \cite{ICDiffuse} gives the effective area for the search,
defined so that the number of detected events is 
\begin{equation}
N = \int dE_\nu d\Omega dt \phi_n(E_{\nu}) A_{eff}(E_\nu).
\label{eq:effarea}
\end{equation}
We use the effective area that is averaged over the entire zenith angle of 90 to 180 degrees (below the horizon), so $\Omega=2\pi$.   We parameterize this effective area with a spline fit, and use Eq. (\ref{eq:effarea}) for a neutrino-flux at the IC40-limit to determine the number of neutrino events $N_{needed}$ required for IceCube to see a signal.   

\begin{figure}[htbp]
\begin{center}
\includegraphics[width=0.5\textwidth]{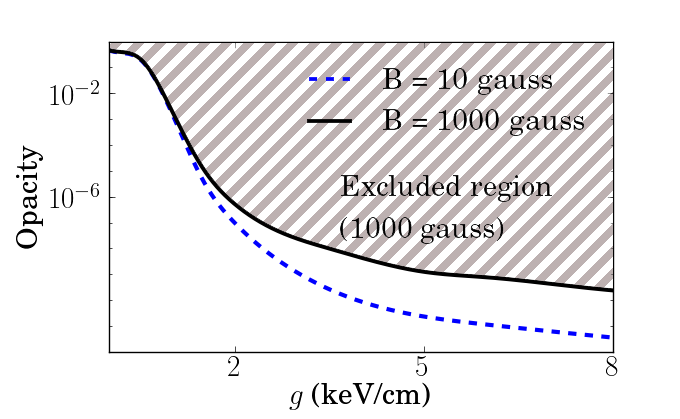}
\caption{Opacity limit as a function of accelerating
  gradient. \label{fig:density}}
\label{fig:limits}
\end{center}
\end{figure}

The IceCube 90\% confidence level limit corresponds to an opacity, $O_{IC} = \phi/\phi_{WB} < 0.4$.  This limit applies at low accelerating gradients; at larger gradients, muon acceleration increases the flux and lowers the allowed opacity. 
Figure\ \ref{fig:limits} shows two dimensional limits on $O$ and $g$.
For $g>5$ keV/cm, the opacity must be below $10^{-8}$, a very severe constraint.   

The opacity limit may be used to set a limit on the matter density in a source; for matter dominated interactions
\begin{equation}
O = \sigma \rho L
\end{equation}
where $\sigma$ is the inelastic interaction cross-section, $\rho$ is the matter density, and $L=E_{p,max}$/g is the length of the accelerator.  We assume that the matter is hydrogen, and take $\sigma = 100$~mb  independent of energy. The actual cross-section rises slowly with energy, but 100\  mb is a reasonable approximation here \cite{LHCSigma}. 
With this, the limit on source density as a function of accelerating gradient is
shown in Fig.\ \ref{fig:density}. 
At very low accelerating gradients, the accelerator gets very long, and 
\begin{equation}
\rho < \frac{O_{IC} g}{\sigma E_{p,{\rm max}}} = 4 \times 10^7 \rm{cm}^{-3}
\frac{g}{1 \rm{ keV/cm}}
\label{eq:density2}
\end{equation}
At very high gradients, the flux is dominated by muon acceleration, and the density limit depends only slightly on the gradient.    Figure \ref{fig:density} shows the density limit as a function of accelerating gradient, for two different magnetic fields.   For accelerating gradients $g> 5$ keV/cm, the source density must be less than $2 \times 10^6$ protons/cm$^3$.   

\begin{figure}[htbp]
\begin{center}
\includegraphics[width=0.5\textwidth]{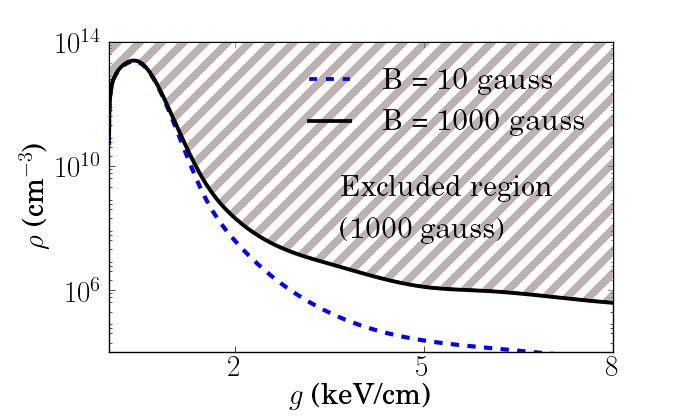}
\caption{Density limit as a function of accelerating
  gradient. \label{fig:density}}
\end{center}
\end{figure}

These limits on $g$, $O$ and $\rho$ can be used to constrain several
models of CR acceleration.    
In the following, we assume that the accelerating gradient and source density are constant.  
The limits would change somewhat by allowing for non-uniform
acceleration, source densities or magnetic fields, or with a different
treatment of the $\pi$ production threshold, 
but the overall conclusions would not change.

Plasma wave acceleration models predict very high accelerating
gradients and at the same time require a significant plasma density,
which is high enough to violate Eq.\ (\ref{eq:density2}).  For
example, in Ref. \cite{Wakefield}, 
a $10^{13}$ keV/cm accelerating gradient requires a density of $\approx 10^{20}$ cm$^{-3}$ in the case of a GRB.   At $10^{13}$ keV/cm, the maximum allowable source density (within the neutrino flux limits) is less than $2 \cdot 10^6$/cm$^3$. In Ref. \cite{MagnetoWakefield}, the same PWA mechanism is applied to an AGN requiring an accelerating gradient of $0.1$ keV/cm and a density of $10^{10}$cm$^{-3}$.   Neither of these models is compatible with 
UHE CR production. 

GRBs have durations of tens of seconds or
less. This requires a very compact  emission region
which argues strongly for a compact
accelerator with high accelerating gradients.  In
the fireball model, shock acceleration occurs in a fireball which is
moving at relativistic speed. CRs are emitted in the direction
of the highly relativistic jets.  
The fireball's Lorentz boosts are estimated to be in the range from a
few, up to about 1,000, see e.g.\ Ref. \cite{grb090510}. Assuming that UHE CR come from
the prompt phase of the GRB, the acceleration must happen
within the $\sim 100$~s of the prompt phase \cite{Vietri1995}. Thus, the accelerating
gradient must be more than ~ $10^{20}$ eV/100 s $\times c$ in the
lab-frame, or about $3\cdot 10^{4}$~keV/cm, so muon and pion acceleration are both
very important.  At an energy of 10 TeV, a $3\cdot 10^{4}$~keV/cm gradient will
raise the neutrino flux by more than 10 orders of magnitude \footnote{At very
large gradients, the pion production threshold needs to be better modeled, but
the enhancement is clearly very large.}.
 
Because GRBs are well localized in time and space, GRB specific
neutrino searches have already set tight constraints, below
the expectations of many calculations \cite{IceCube_GRB_Nature}.
Improved modeling may lead to somewhat lower fluxes, see
e.g.\ \cite{Winter02}.   However, accelerator gradients
larger than 5 keV/cm would certainly have raised the neutrino flux to
the point where it would be easily detectable by IceCube, at least for models where the neutrinos come from pion decay.
The neutrino flux limits appear inconsistent with current fireball phenomenology and the large accelerating
gradients imposed by the length of the prompt phase. 

Magnetars are newly formed neutron stars with magnetic fields up to $10^{15}$~Gauss.
The acceleration of CRs may occur
 in regions with large electric fields, but low magnetic
fields \cite{Arons2003}.  The limited size of these regions ($\sim$
30,000 km) require  gradients of  $3\cdot 10^{7}$ keV/cm.   If magnetars are the primary
sources of UHE protons,
$10^{18}$ eV muons are produced. At a gradient of
$3\cdot 10^{7}$~keV/cm, muon acceleration would greatly magnify the neutrino flux. This 
precludes magnetars as being the dominant source of UHE cosmic-rays
unless the matter density in the source is very low, less than
$2 \cdot 10^{6}$/cm$^3$.

Finally, if CRs are mostly heavier nuclei, then muon acceleration has
an even larger effect.  Since the CR energy is divided among
many $A$ nucleons, the maximum neutrino energy is reduced to
$0.08/A$ of the maximum CR energy, and so the neutrino flux at high energies is heavily suppressed.  However, if the muons are significantly accelerated before they decay, then the maximum energy increases to 30\% - a very large increase. 

In conclusion, many CR acceleration models require high accelerating
gradients that lead to significant muon acceleration.  For
gradients above 1.6 keV/cm, the neutrino production rate
is enhanced significantly in CR sources and the maximum neutrino energy increases. For photohadronic interactions, the maximum energy rises from about 8\% of the primary proton energy to up to about 30\% of the maximum attainable energy. For matter dominated sources, this may be higher than the maximum proton energy.  
Current IceCube neutrino flux limits have been used to set
two-dimensional limits on accelerating gradient and source opacity or
target density.  When muon acceleration is considered, the IceCube
limits exclude several current models for UHE cosmic-ray acceleration,
such as plasma wave acceleration at different types of sources, the
fireball model of GRBs, and magnetars, unless the particle density in
the objects is extremely low.

We thank R.\ Schlickeiser for discussions.  This work was funded in part by the U.S. National Science Foundation under grant 0653266 and the U.S. Department of Energy under contract number DE-AC02-05CH11231.

\end{document}